\documentclass[twocolumn,english,aps,prl,superscriptaddress,showpacs]{revtex4}
\usepackage[T1]{fontenc}
\usepackage{amsmath}
\usepackage{graphicx}
\usepackage{amssymb}
\usepackage{epsfig}

\usepackage{babel}
\usepackage{dsfont}

\usepackage{babel}

\begin{document}

\title{Inverse design of cooperative electromagnetic interactions}

\author{M. Langlais}
\affiliation{Laboratoire Charles Fabry,UMR 8501, Institut d'Optique, CNRS, Universit\'{e} Paris-Sud 11,
2, Avenue Augustin Fresnel, 91127 Palaiseau Cedex, France.}
\affiliation{Total New Energies, R\&D, Tour Michelet, 92078 Paris La D\'{e}fense Cedex, France.}
\author{J. P. Hugonin}
\affiliation{Laboratoire Charles Fabry,UMR 8501, Institut d'Optique, CNRS, Universit\'{e} Paris-Sud 11,
2, Avenue Augustin Fresnel, 91127 Palaiseau Cedex, France.}
\author{M. Besbes}
\affiliation{Laboratoire Charles Fabry,UMR 8501, Institut d'Optique, CNRS, Universit\'{e} Paris-Sud 11,
2, Avenue Augustin Fresnel, 91127 Palaiseau Cedex, France.}
\author{P. Ben-Abdallah}
\email{pba@institutoptique.fr}
\affiliation{Laboratoire Charles Fabry,UMR 8501, Institut d'Optique, CNRS, Universit\'{e} Paris-Sud 11,
2, Avenue Augustin Fresnel, 91127 Palaiseau Cedex, France.}

\date{\today}


\begin{abstract}

The cooperative electromagnetic interactions between discrete resonators have been widely used to modify the optical properties of metamaterials. Here we propose a general evolutionary approach for engineering these interactions in arbitrary networks  of resonators. To illustrate the performances of this approach, we designed by genetic algorithm, an almost perfect broadband absorber in the visible range made with a simple binary array of metallic nanoparticles.

\end{abstract}

\maketitle

Engineering light-matter interactions is a longstanding problem in physics and is of prime importance for numerous technological applications such as the photo and thermophovoltaic energy conversion~\cite{Atwater, Bermel}, the optical manipulation of nanoobjects \cite{Ashkin} or the quantum information treatment \cite{Poddubny}. Light interaction with resonant structures embedded inside a material is a natural way to modify its optical properties. To date, a large number of resonant structures have been developped following such a strategy. Among these, metamaterials based on metallo-dielectric structures have been proposed to operate at frequencies ranging from the microwave domain \cite{Padilla} to the visible \cite{Atwater2}. 

The design of artificially constructed magnetodielectric resonators which strongly interact cooperatively is a very recent and promising way to generate metamaterials that highlight innovative physical \cite{Jenkins,Fedotov} and transport \cite{Ben-Abdallah3} properties. However, so far, only heuristic approaches have been followed to identify the convenient meta-structures which display target functionalities. 
In the present Letter, we present a general theory to describe the multiple scattering interactions mechanisms in discret  networks of resonators embedded in a host material and we propose an evolutionary method  to identify the appropriate inner structure of  networks that highlight a targeted optical property.  To illustrate the strong potential of cooperative interactions to tailor the optical properties of materials we design, by using a genetic algorithm, a broadband light absorber made with simple binary lattices of metallic nanoparticles immersed in a transparent host material. 

To start, let us consider a set of objects dispersed inside a host material as depicted in Fig. 1. Suppose this system is higlighted by an external harmonic electromagnetic field of wavelength much larger than the typical size of objects. In this condition, we can associate to each object an electric (E) and magnetic (H) dipole moment $\mathbf p_{i;A} (A=E,H)$ (the higher orders contributions are discussed in the supplemental material \cite{Supplemental}). The local electromagnetic field $\mathbf{A}^{ext}(\mathbf{r}_i)$ at the dipoles location $\mathbf{r}_i$ results from the superposition of external incident field, the field generated by the others dipoles and the auto-induced field which comes from  the interactions with the interfaces. Therefore it takes the form
\begin {equation}
 \mathbf A^{ext}_i=\mathbf A^{inc}_i-i\omega \underset{B=E,H}{\sum}\Gamma_{AB}(\underset{j\neq i}{\sum}\mathds{G}_{ij}^{AB}\mathbf p_{j;B}+\Delta \mathds{G}_{ii}^{AB}\mathbf p_{i;B}),\label{Eq:external_field}
\end{equation}
where $\left(\begin{array}{cc}
\Gamma_{EE} & \Gamma_{EH}\\
\Gamma_{HE} & \Gamma_{HH}
\end{array}\right)\equiv\left(\begin{array}{cc}
i\omega\mu_{0} & \omega/c\\
\omega/c & -i\omega\varepsilon_{0}
\end{array}\right)$ and $\mathds{G}_{ij}^{AB}$ is the dyadic Green tensor in the host material which takes into account the presence of interfaces and gives the field $\mathbf A$ at the position $\mathbf r_i$ given a $\mathbf B$-dipole located in $\mathbf r_j$. $\Delta \mathds{G}^{AB}$ defined as $\Delta \mathds{G}^{AB}\equiv \mathds{G}^{AB}-\mathds{G}^{AB}_0$  
gives the contribution of interfaces only. Here $\mathds{G}^{AB}_{\mathrm{0}}(\mathbf{r}_i,\mathbf{r}_j)=\frac{\exp({\rm i}kr_{ij})}{4\pi r_{ij}}\times\begin{cases}
\left[\left(1+\frac{{\rm i}kr_{ij}-1}{k^{2}r_{ij}^{2}}\right)\mathds{1}+\frac{3-3{\rm i}kr_{ij}-k^{2}r_{ij}^{2}}{k^{2}r_{ij}^{2}}\widehat{\mathbf{r}}_{ij}\otimes\widehat{\mathbf{r}}_{ij}\right]\:if A=B\\
\frac{{\rm i}kr_{ij}-1}{k r_{ij}}\mathds{L} \:\:\:\:\:\:\:\:\:\:\:\:\:\:\:\:\:\:\:\:\:\:\:\:\:\:\:\:\:\:\:\:\:\:\:\:\:\:\:\:\:\:\:\:\:\:\:\:\:\:\:\:\:\:\:\:\:\:\:\:\:\:\:\:\:\:\:\:if A\neq B
\end{cases}$
is the free space Green tensor in the host material defined with the unit vector $\widehat{\mathbf{r}}_{ij}\equiv\mathbf{r_{\mathit{ij}}}/r_{ij}$.
$\mathbf{r_{\mathit{ij}}}$ denotes here the vector linking the center of
dipoles i and j, while $r_{ij}=\mid\mathbf{r}_{ij}\mid$, $k$ is the wavector,$\mathds{1}$
the unit dyadic tensor and $\mathds{L}=\left(\begin{array}{ccc}
0 & -\hat{r}_{ij,z} & \hat{r}_{ij,y}\\
\hat{r}_{ij,z} & 0 & -\hat{r}_{ij,x}\\
-\hat{r}_{ij,y} & \hat{r}_{ij,x} & 0
\end{array}\right)$. 
Beside the dipoles location the auto-induced part of field does not exist anymore and it takes the simplified form
\begin {equation}
 \mathbf A^{ext}(\mathbf{r})=\mathbf A^{inc}(\mathbf{r})-i\omega \underset{B=E,H}{\sum}\Gamma_{AB}\underset{j}{\sum}\mathds{G}^{AB}(\mathbf{r}-\mathbf{r}_j)\mathbf p_{j;B}.\label{Eq:external_field2}
\end{equation}

\begin{figure}[Hhbt]
\includegraphics[scale=0.3]{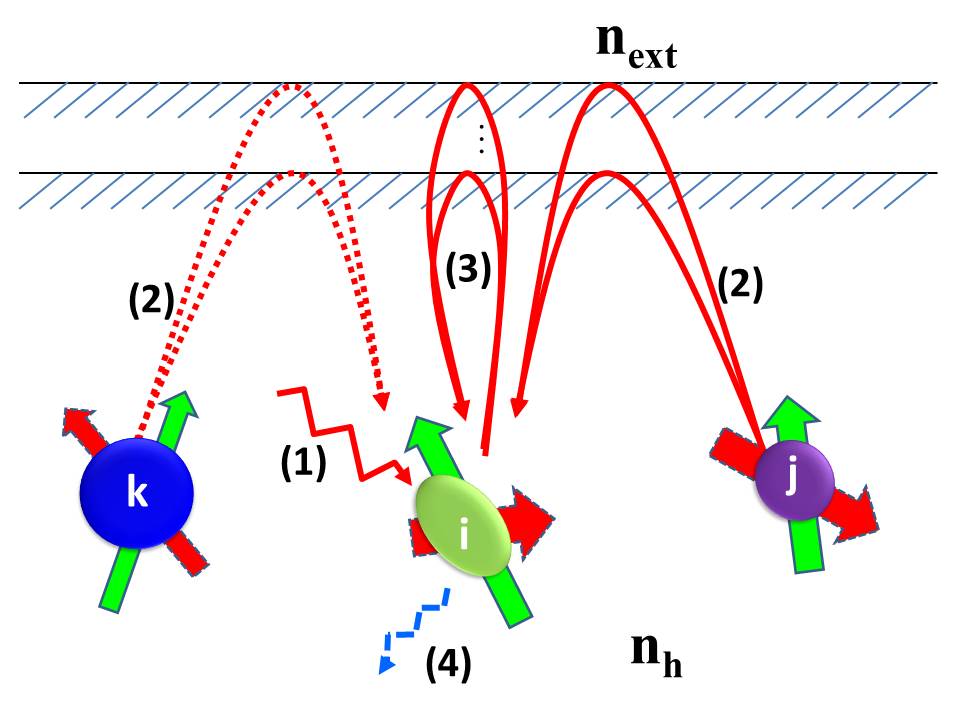}
\caption{Multiple light scattering interactions in a set of subwavelength plasmonic structures embeded in a transparent host material of refractive index $n_h$. In the dipolar approximation each object is replaced by both a dipolar electric moment and a magnetic moment. The external field felt by each object decomposes into (1) the incident field, (2) the field radiated by the other objects and (3) the auto-induced field which comes from the interface after being emitted by the object itself. All dipoles radiate (4) in their surrounding. } 
\end{figure}

It immediately follows that, the dipolar moments associated to each object reads
\begin {equation}
\mathbf p_{i;A}=\chi_A \mathbf{\overleftrightarrow{\alpha}}_{i;A}\mathbf A^{ext}_i\label{Eq:dipole_moment}
\end{equation}
where $\chi_A$ represents either the vacuum permittivity $\varepsilon_{0}$ or the vacuum permeability $\mu_{0}$ and $\mathbf{\overleftrightarrow{\alpha}}_{i,A}$ is the free polarizability tensor of $i^{th}$ object under the action of field $\mathbf A$. By inserting the external contribution (\ref{Eq:external_field}) of  local field into relation (\ref{Eq:dipole_moment}) we get the following system which relates all dipole moments
\begin {equation}
\mathbf p_{i;A}=\chi_A \mathbf{\overleftrightarrow{\alpha}}_{i;A}[\mathbf A^{inc}_i-i\omega\underset{j}{\sum} \underset{B=E,H}{\sum}\mathds{G}_{reg}^{AB}(r_i,r_j)\mathbf p_{j;B}].\label{Eq:dipole_system}
\end{equation}
Here, where we have introduced the regularized Green tensor
\begin {equation}
\mathds{G}_{reg}^{AB}(r,r^{'})=\begin{cases}
\quad\Gamma_{AB}\mathds{G}^{AB}(r,r^{'}) \; if \; r\neq r^{'}\\
\Gamma_{AB}\Delta \mathds{G}^{AB}(r,r^{'})\; if \;r= r^{'}
\end{cases}.\label{Eq:Green_reg}
\end{equation}

In the particular case of n-ary periodic lattices made with $n$ arbitrary dipoles of free electric or magnetic polarizability $\overleftrightarrow{\alpha}_{i;A=E,H}$ distributed in a unit cell we have, according to the periodicity, the supplementary relations for the incident fields $\mathbf A_{j_{\beta}}^{inc}=\widetilde{\mathbf A_{\beta}}exp(i \mathbf k_{//}.\mathbf r_{j_{\beta}})$
and  for the dipolar moments $\mathbf p_{j_{\beta};A}=\widetilde{\mathbf p}_{{\beta};A}exp(i \mathbf k_{//}.\mathbf r_{j_{\beta}})$. Here $\mathbf r_{j_{\beta}}$ is the position vector of the $\beta^{th}$ dipole inside the unit cell $j$ of lattice and $\mathbf k_{//}$ is the parallel component of wavector. 
\begin{figure}[Hhbt]
\includegraphics[scale=0.33]{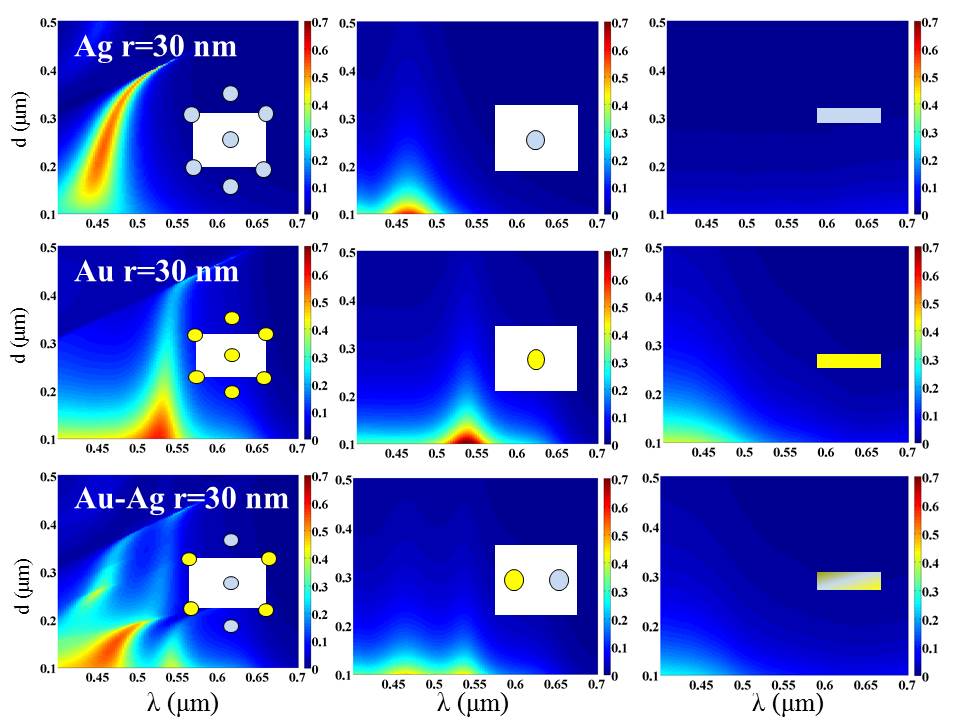}
\caption{On the first column, absorption of simple and binary hexagonal lattices made with Ag and Au nanoparticles 30 nm radius immersed at $h=100 nm$ from the surface in a transparent host medium of index  $n_h=1.5$ with respect to the density in particles. On the second column, this absoption is compared with the absorption of single particles without multiple scattering interaction and, on the last column, with the results given by the effective medium theory with the same filling factor.} 
\end{figure} 
Accordingly, Eq. (\ref{Eq:dipole_system}) can be solve with respect to the incident field to give 
\begin {equation}
\left(\begin{array}{c}
\widetilde{\mathbf p}_{E}\\
\widetilde{\mathbf p}_{H}
\end{array}\right)=\mathcal{A}^{-1}\mathcal{H}
\left(\begin{array}{c}
\widetilde{\mathbf E}\\
\widetilde{\mathbf H}
\end{array}\right).\label{Eq:dress_polarizability}
\end{equation}
Here we have set $\widetilde{\mathbf p}_{A=E,H}=\left(\begin{array}{ccc}
\widetilde{\mathbf p}_{1,A},&\cdots,&\widetilde{\mathbf p}_{n,A}
\end{array}\right)^{t}$ and $\widetilde{\mathbf A}=\left(\begin{array}{ccc}
\widetilde{\mathbf A}_1,&\cdots,&\widetilde{\mathbf A}_n
\end{array}\right)^{t}$ 
and we have define the block matrixes
\begin {equation}
\mathcal{H}=diag(\varepsilon_{0}\overleftrightarrow{\alpha}_{1;E},\ldots,\varepsilon_{0}\overleftrightarrow{\alpha}_{n;E},\mu_{0}\overleftrightarrow{\alpha}_{1;H},\ldots,\mu_{0}\overleftrightarrow{\alpha}_{n;H})\label{Eq:polarizability}
\end {equation}
and
\begin{widetext}
\begin {equation}
\mathcal{A}=
\left(\begin{array}{ccccccc}
(\mathds{1}+\mathds{U}_{11}^{EE}) & \mathds{U}_{12}^{EE} & \cdots & \mathds{U}_{1n}^{EE} & \mathds{U}_{11}^{EH} & \cdots & \mathds{U}_{1n}^{EH}\\
\mathds{U}_{21}^{EE} & \ddots &  & \vdots & \vdots &  & \vdots\\
\vdots & \ddots & (\mathds{1}+\mathds{U}_{nn}^{EE}) & \mathds{U}_{n-1,n}^{EE} & \mathds{U}_{n1}^{EH} & \cdots & \mathds{U}_{nn}^{EH}\\
\mathds{U}_{n1}^{EE} & \cdots & \mathds{U}_{n,n-1}^{EE} & (\mathds{1}+\mathds{V}_{11}^{HH}) & \mathds{V}_{12}^{HH} & \cdots & \mathds{V}_{1n}^{HH}\\
\mathds{V}_{11}^{HE} & \cdots & \mathds{V}_{1n}^{HE} & \mathds{V}_{21}^{HH} & \ddots & \ddots & \vdots\\
\vdots &  & \vdots & \vdots & \ddots & \ddots & \mathds{V}_{n-1,n}^{HH}\\
\mathds{V}_{n1}^{HE} & \cdots & \mathds{V}_{nn}^{HE} & \mathds{V}_{n1}^{HH} & \cdots & \mathds{V}_{n,n-1}^{HH} & (\mathds{1}+\mathds{V}_{nn}^{HH})
\end{array}\right)
\end {equation}
\end{widetext}
with
\begin {equation}
\mathds{U}_{lk}^{EA}=i\varepsilon_{0}\omega\overleftrightarrow{\alpha}_{l;E}\underset{j}{\sum}\mathds{G}_{reg}^{EA}(\mathbf {r}_{0l},\mathbf{r}_{jk})e^{i\mathbf k_{//}.(\mathbf {r}_{jk}-\mathbf {r}_{0l})},
\end {equation}
\begin {equation}
\mathds{V}_{lk}^{HA}=i\mu_{0}\omega\overleftrightarrow{\alpha}_{l;H}\underset{j}{\sum}\mathds{G}_{reg}^{HA}(\mathbf {r}_{0l},\mathbf{r}_{jk})e^{i\mathbf k_{//}.(\mathbf {r}_{jk}-\mathbf {r}_{0l})}.
\end {equation}
Relation (\ref{Eq:dress_polarizability}) defines the dressed polarizability tensor
\begin {equation} 
\mathbf\Upsilon\equiv\mathcal{A}^{-1}\mathcal{H}=
\left(\begin{array}{cc}
\mathbf\Upsilon^{EE} & \mathbf\Upsilon^{EH}\\
\mathbf\Upsilon^{HE} &\mathbf\Upsilon^{HH}
\end{array}\right)
\end {equation}
of resonators within the unit cell of lattice. We clearly see that this polarizability is resonant when $det \mathcal{A}=0$. The former condition depends only on the intrinsic properties of materials while the second is of geometric nature and depends on the spatial distribution of nanostructures \cite{Ben-Abdallah2}.

The power dissipated inside each resonator at a frequency $\omega$ is given by the rate of doing work by the fields in its volume $V_i$ 
\begin {equation}
 \mathcal{P}_i(\omega)=\frac{1}{2}\underset{A=E,H}{\sum}\underset{V_i}{\int}Re[\mathbf j^*_{A}(\mathbf r,\omega).\mathbf A(\mathbf r,\omega)]d\mathbf r.\label{Eq:power}
\end{equation}
 Here $\mathbf A$ denotes either the local electric or magnetic field $\mathbf E$ and  $\mathbf H$ while $\mathbf j_{E}$ and  $\mathbf j_{H}$  are the corresponding local current density. In the dipolar approximation $\mathbf j_{i;A}=-i\omega \mathbf p_{i;A}\delta(\mathbf r-\mathbf r_i)$, expression (\ref{Eq:power}) can be recasted into the discrete form
\begin {equation}
 \begin{split}\mathcal{P}_i(\omega)=-\frac{\omega }{2}\underset{A=E,H}{\sum}  \{Im[\mathbf p^*_{i;A}(\omega).\mathbf A_i^{ext}(\omega)]\\
-\frac{\omega^3\mu_0}{ 2}\mathbf p^*_{i;A}Im[\mathds{G}^{AA}_{\mathrm{0}}(\mathbf r_i,\mathbf r_i)]\mathbf p_{i;A}\}.
\end{split} \label{Eq:power2}
\end{equation}
By inverting (\ref{Eq:external_field}) after having replaced the dipole moments by their expression with respect to $\mathbf A_i^{ext}$, we can express $\mathbf A_i^{ext}$ in term of $\mathbf A_{inc}$
and explicitely calculate the power dissipated in each object under an external lighting.

For spherical particles of radius $R$ the polarizability is straightforwardly derived from the Mie scattering theory \cite{Bohren}. If those particles, of refractive index $n_m$, are immersed inside a medium of index $n_h$, we have $\overleftrightarrow{\alpha}_{A}={\alpha}_{A}\mathds{1}$ with
\begin {equation} 
{\alpha_E}^{-1}=k^3_0\frac{n_h}{6\pi}(C_E-i),
\end {equation}
\begin {equation} 
{\alpha_H}^{-1}=k^3_0\frac{n^3_h}{6\pi}(C_H-i).
\end {equation}
Here
\begin{widetext}
\begin {equation} 
C_{E}=\frac{\frac{\rho_{m}^{2}-\rho_{h}^{2}}{\rho_{m}^{2}\rho_{h}^{2}}(Cos\rho_{h}+\rho_{h}Sin\rho_{h})(Sin\rho_{m}-\rho_{m}Cos\rho_{m})+\rho_{m}Cos\rho_{h}Cos\rho_{m}+\rho_{h}Sin\rho_{h}Sin\rho_{m}}{\frac{\rho_{h}^{2}-\rho_{m}^{2}}{\rho_{m}^{2}\rho_{h}^{2}}(Sin\rho_{h}-\rho_{h}Cos\rho_{h})(Sin\rho_{m}-\rho_{m}Cos\rho_{m})-\rho_{m}Sin\rho_{h}Cos\rho_{m}+\rho_{h}Cos\rho_{h}Sin\rho_{m}}\label{Eq:cross1}
\end {equation}
and
\begin {equation} 
C_{H}=\frac{-\rho_{h}^{2}Cos\rho_{h}(Sin\rho_{m}-\rho_{m}Cos\rho_{m})+\rho_{m}^{2}Sin\rho_{m}(Cos\rho_{h}+\rho_{h}Sin\rho_{h})}{\rho_{h}^{2}Sin\rho_{h}(Sin\rho_{m}-\rho_{m}Cos\rho_{m})-\rho_{m}^{2}Sin\rho_{m}(Sin\rho_{h}-\rho_{h}Cos\rho_{h})}\label{Eq:cross2}
\end {equation}
\end{widetext}
with $\rho_{h}=k_0 n_h R$ and $\rho_{m}=k_0 n_m R$. According to  Eqs. (\ref{Eq:power2}), (\ref{Eq:cross1}) and (\ref{Eq:cross2}) it follows that the power disipated in each particle can be expressed both  in term of absorption cross-sections and of incident external field
\begin {equation}
  \begin{split}\mathcal{P}_i(\omega)=-\frac{\omega}{2}\{\varepsilon_{0}\frac{n_{h}\omega^{3}}{6\pi c^{3}}Im[\mathbf E_{i}^{ext*}(C_{E} \overleftrightarrow{\alpha}^*_{E,i} \overleftrightarrow{\alpha}_{E,i})\mathbf E_{i}^{ext}] \\
+\mu_{0}\frac{n^3_{h}\omega^{3}}{6\pi c^{3}}Im[\mathbf H_{i}^{ext*}(C_{H} \overleftrightarrow{\alpha}^*_{H,i} \overleftrightarrow{\alpha}_{H,i})\mathbf H_{i}^{ext}]\}
\end{split} \label{Eq:power3}
\end{equation}

To illustrate this, we show in  Fig. 2 the absorption spectra at normal incidence in the visible range for TE waves of single Au and Ag spherical particles \cite{Palik} dispersed in regular hexagonal lattices of side length $d$ and for binary Au-Ag  lattices. All lattices are  immersed in a transparent material of refractive index $n_h=1.5$ and are maintained at a distance $h=100 nm$ from the surface.  We clearly see by comparing these absorption spectra with those of isolated particles that the resonance peaks in single particle lattices are essentialy centered at the resonance frequency of free particles. However the cooperative interactions allow to increase the absorption level even in diluted lattices where the filling factor $f$ i below $3\%$. This can be explained by comparing this absorption with the absorption of isolated particles (Fig. 2). It clearly appears that the absorption cross-section of particles is strongly enhanced by the presence of neighborhood particles. The comparison of the overall absorption of nanoparticle lattices with that of simple metallic films of thickness defined from the filling factors points out importance of cooperative effects in  the absorption of  lattices. In binary lattices, new configurationnal resonances add up to the resonances of single lattices and naturally enlarge the absorption spectrum. 
\begin{figure}[Hhbt]
\includegraphics[scale=0.35]{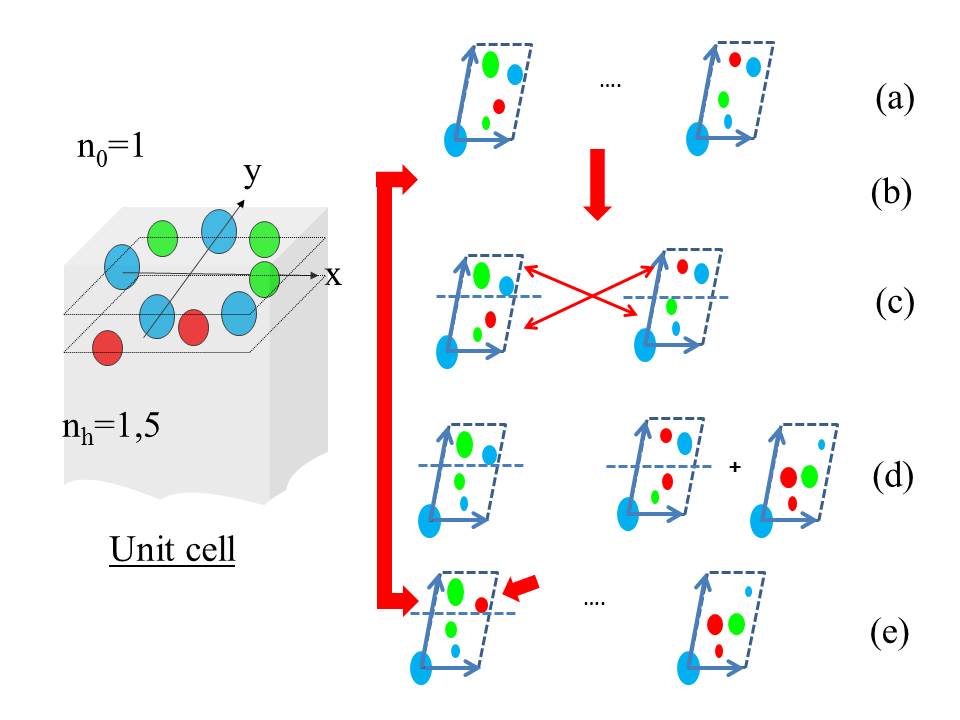}
\caption{Evolutionary algorithm to optimize a n-ary lattice. (a) A random population of periodic lattices (a physical view of an unit cell is plotted on the left) is randomly generated. (b) The best individus basd on the fitness function are selected as parents for the crossing over. (c) The next generation is created by linear crossing and completed by new individus (d) to keep the total population constant. (e) Mutations are aaplied on a few number of individus (typically 5\%) in the current generation.} 
\end{figure} 

In the following, we show the strong potential of cooperative interactions mechanisms between resonators to tailor the optical properties of materials. For this purpose we present the inverse design of a broadband absorber in the visible range  \cite {Narimanov,Pendry,Atwater2,Sergeant} made with  a n-ary array metallic spherical nanoparticles. A n-ary lattice is defined from a unit cell $\mathcal C$ of a two dimensional paving with a certain thickness (see Fig. 3). In the unit cell of a lattice we consider a set of $n$ vectors $\mathbf r_i$ and $n$ positive reals $R_i$ that represent the location of particles center and the radius of particles, respectively. To avoid the particle interpenetration  these vectors must satisfy to the supplemental constraint $\mid \mathbf r_i-\mathbf r_j\mid>R_i+R_j$ . To design the n-ary lattice in order to maximize its overalll absorption we have to explore the large and complex space of all possible configurations. To do that we employ a genetic algorithm (GA) \cite{Holland} which is a stochastic global optimisation method that is based on natural selection rules in a similar way to the Darwin's theory of evolution.  Evolutionary optimization has been yet successfully applied in numerous fields of optics \cite{Feichtner, Ben-Abdallah1}. 
\begin{figure}[Hhbt]
\includegraphics[scale=0.32]{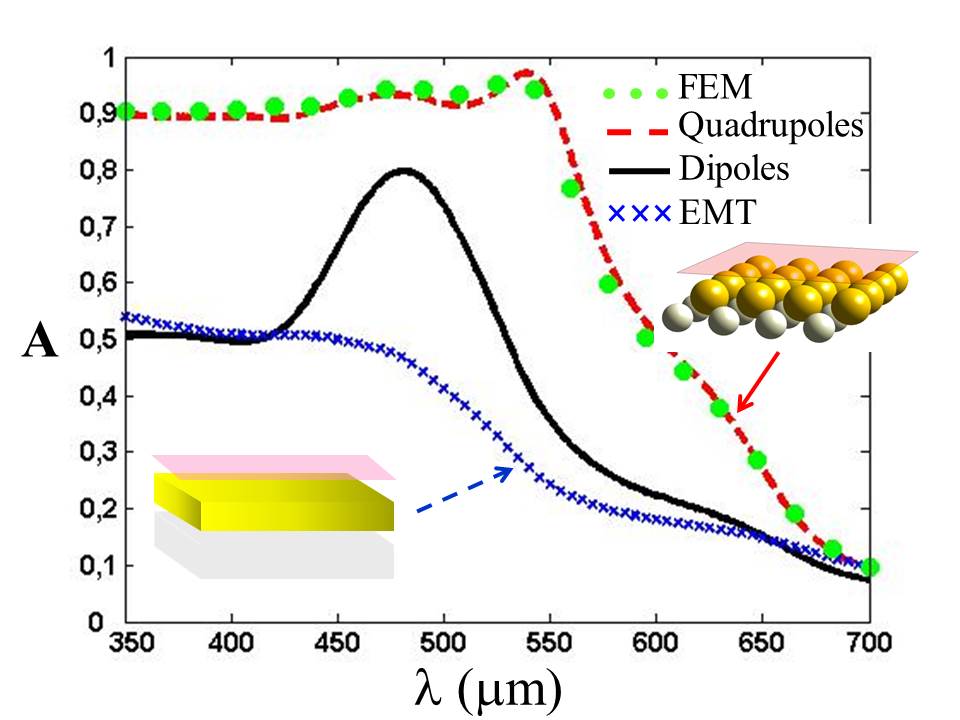}
\caption{Light absorption spectrum at normal incidence of a binary Au-Ag lattice (blue dashed curve) optimzed by GA by taking into account all multipolar interactions until the second  order (quadrupoles) and of a multilayer based on Au-Ag films of thickness defined with the filling factor in nanoparticles (i.e. effective medium theory). Circles curve shows the result obtained by solving the Maxwell's equations with a finite element method.} 
\end{figure} 
Basically, a GA uses an initial population (Fig. 3) of typically few hundreds ofstructures also called individus which are randomly generated in size and position. For each  individus we calculate the fitness parameter which is here the mean absorption (for a given polarization) $\overline{A}=(\lambda_{max}-\lambda_{min})^{-1}\intop_{\lambda_{min}}^{\lambda_{max}}A(\lambda)d\lambda$ over the spectral range $[\lambda_{min};\lambda_{max}]$ where we want to increase the absorption. The monochromatic absorption $A(\lambda)$ at a given wavelength is simply given by the the sum of power dissipated inside the particles of the unit cell normalized by the incident flux $\phi_{inc}(\lambda)$ on its surface $\mathfrak{A}$ that is
\begin {equation} 
A(\lambda)=\frac{\underset{i\in Cell}{\sum}\mathcal{P}_i(\lambda)}{\mathfrak{A}\phi_{inc}(\lambda)}.\label{Eq:absorption}
\end {equation}
\begin{figure}[Hhbt]
\includegraphics[scale=0.35]{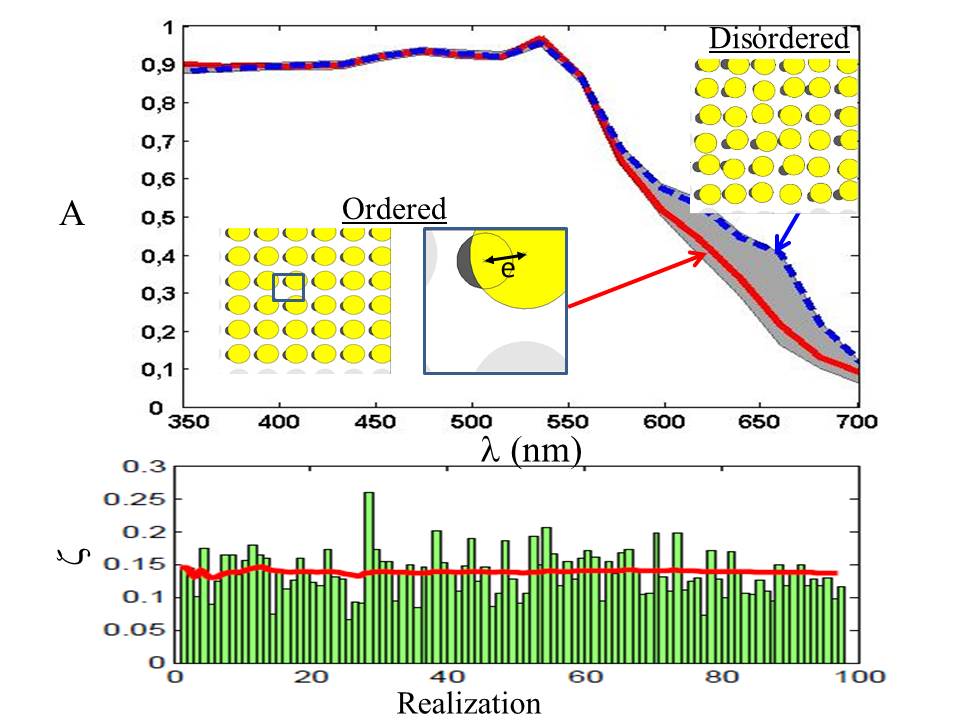}
\caption{Impact of disorder on the light absorption spectrum at normal incidence in a binary Au-Ag lattice.The spatial location of particles is randomly perturbated by a displacement of $20 nm$. The red ciurve corresponds to the spectrum (in polarization TM at nomrla incidence) of the optimized structure and the dotted blue curve is the spectrum of a particular random realization (results in polarization TE, not plotted here are similar). The dashed area shows the maximum and minimm values of absorption spectrum of different random realizations. The  histogram shows the discrepancy with the optimal fintess for different realizations of the structure. The red line on the histogram shows the mean error with respect to the number of realizations.} 
\end{figure} 
The GA consists in maximizing the fitness function of structures (i.e. $\overline{A}\rightarrow max$). To do so, we select 90\% of the highest fitness as future parents for the next generation of selecting process. Those parents are linear crossed and the new 'child' generation is completed by new individual structures (randomly generated) to keep the same total number of lattices for any generation. To avoid the convergence toward local extrema,  every m (typically 10) generations,  we introduce also some mutations that is random perturbations with a probability of about 5\% on the value of parameters we are optimizing.  The results presented in Fig. 4  for supperposed binary Au-Ag lattices (with the radius $r_{Au}=77 nm$ and $r_{Ag}=39 nm$, the separation distances from the surface $h_{Au}=120 nm$ and $h_{Ag}=242 nm$, the lattice constants  $d_{Au}=d_{Ag}=200 nm$ and the off-centring $e_x=56 nm$ and $e_y=10 nm$ ) exhibit a broad absorption band in the visible range. By taking into account the multipolar interactions until the second order (i.e. quadrupolar interactions) we see that the level of aborption becomes close to one (the comparaison of these results with full electromagnetic simulations shows that the highest order multipole moments  do not contribute significantly to the overall absorption). 

Interestingly, the numerical simulations have shown  also that the cooperative effects are not too much sensitive to the presence of disorder. In Fig. 5 we see that, by disturbing the optimal structure with a random perturbation of particles locations by a maximum displacement of $20 nm$, the discrepancy between the optimal structure and the perturbed ones, given by the mean square error $\zeta=[\intop_{\lambda_{min}}^{\lambda_{max}}(A(\lambda)-A_{opt}(\lambda))^2d\lambda]^{1/2}$, remains small. 

In conclusion, our results have demonstrated the strong potential of inverse design of cooperative interactions between optical resonators. We believe that this approach opens the way  for a rational design of metamaterials and it could find broad applications in various fields of photonics such as solar cells, quantum information systems and, according to the reciprocity principle \cite{Landau}  light extraction technologies. 

%
%

\begin{acknowledgments}
J.-P. H. acknowledges discussions with  J.J. Greffet. P. B.-A. gratefully acknowledges the support of Total news energies.
\end{acknowledgments}


\begin{thebibliography}{26}

\bibitem{Atwater} H. A. Atwater and A. Polman, Nature Materials, {\bf 9}, 205-2013 (2010). 
\bibitem{Bermel} P. Bermel et al., Opt. Express {\bf 18}, A314-A334 (2010). 
\bibitem{Ashkin} Ashkin A., Dziedzic J. M., Bjorkholm J. E. and Chu  S.,  Observation of a single-beam gradient force optical trap for dielectric particles, Opt. Lett., {\bf 11}, 288–290 (1986).
\bibitem{Poddubny} A. N. Poddubny, P. A. Belov and Y. S. Kivshar, Phys. Rev. A, {\bf84}, 023807 (2011).
\bibitem{Padilla} N. I. Landy, S. Sajuyigbe, J. J. Mock, D. R. Smith and W. J. Padilla, Phys. Rev. Lett. {\bf 100}, 207402 (2008).
\bibitem{Atwater2} K. Aydin, V. E. Ferry, R. M. Briggs and H. A. Atwater, Nat. Comms. {\bf 2}, 517 (2011).
\bibitem{Jenkins} S. D. Jenkins and J. Ruostekoski,Phys. Rev. Lett., {\bf 111}, 147401 (2013).
\bibitem{Fedotov} V. A. Fedotovi, N. Papasimakis, E. Plum, A. Bitzer M. Walther, P. Kuo, D. P. Tsai and N. I. Zheludev, Phys. Rev. Lett., {\bf 1104}, 1223901 (2010).
\bibitem{Ben-Abdallah3} P. Ben-Abdallah, R. Messina, S.-A. Biehs, M. Tschikin, K. Joulain, and C. Henkel, Phys. Rev. Lett., {\bf 111}, 174301 (2013). 
\bibitem{Supplemental} See Supplemental Material at (added by Editors).
\bibitem{Ben-Abdallah2} P. Ben-Abdallah, S.-A. Biehs, and K. Joulain, Phys. Rev. Lett. {\bf 107}, 114301 (2011). 
\bibitem{Bohren} C. F. Bohren, D. R. Huffman, Absorption and Scattering of Light by Small Particles (Wiley Science, New York, 1998).
\bibitem{Narimanov} E. E. Narimanov, A. V. Kildishev, Appl. Phys. Lett. 95, 041106 (2009).
\bibitem{Pendry} A. Aubry et al., Nano. Lett. 10. 2574-2579 (2010).
\bibitem{Sergeant} N. P. Sergeant, O. Pincon, M. Agrawal,  P. Peumans, Optics Express {\bf 17}, 25, pp.22800-22812 (2009).
\bibitem{Holland} J. H. Holland, Adaptation in Natural and Artificial Systems (MIT Press/Bradford Books Edition, Cambridge, MA, 1992). 
\bibitem{Feichtner} T. Feichtner, O. Selig, M. Kiunke, and B. Hecht, Phys. Rev. Lett. {\bf 109}, 127701 (2012). 
\bibitem{Ben-Abdallah1} J. Drevillon and P. Ben-Abdallah, J. Appl. Phys. {\bf 102}, 114305 (2007).
\bibitem{Palik} E. D. Palik, Handbook of Optical Constants of Solids (Academic Press, New York, 1998).
\bibitem{Landau} L. Landau, E. Lifchitz, and L. Pitaevskii, Electromagnetics of Continuous Media (Pergamon, Oxford, 1984).











\end{thebibliography}
\end{document}